\documentclass[floatfix,reprint, pre, cha, twocolumn, superscriptaddress]{revtex4-1}

\usepackage{graphicx}
\usepackage{epstopdf}

\usepackage{dcolumn}

\usepackage[right]{lineno}

\usepackage{bm}
\usepackage{amssymb}
\usepackage{amsmath}
\usepackage{bbold}

\usepackage{hyperref} 

\newcommand{\um}[1]{{\"#1}}

\nolinenumbers

\newcommand{\COV}{\ensuremath{\sigma}}
\newcommand{\CORR}{\ensuremath{C}}
\newcommand{\AUC}{\ensuremath{\text{AUC}}}
\newcommand{\orderof}[1]{\ensuremath{\mathcal{O}(#1)}} 
\newcommand{\dd}{\ensuremath{\text{d}}}
\newcommand{\unity}{\ensuremath{\mathbb{1}}}
\newcommand{\T}{\ensuremath{{^\intercal}}}

\DeclareMathOperator{\Mod}{mod}

\newlength{\figwidth}
\begin{document}

\title{Transition to reconstructibility in weakly coupled networks}
\author{Benedict L\um{u}nsmann}
\affiliation{Network Dynamics, Max Planck Institute for Dynamics and Self-Organization (MPIDS), 37077 G\um{o}ttingen, Germany}
\affiliation{Max Planck Institute for the Physics of Complex Systems (MPIPKS), 01187 Dresden, Germany}

\author{Christoph Kirst}
\affiliation{Network Dynamics, Max Planck Institute for Dynamics and Self-Organization (MPIDS), 37077 G\um{o}ttingen, Germany}
\affiliation{Rockefeller University,  NY 10065-6399 New York, USA}

\author{Marc Timme}
\affiliation{Network Dynamics, Max Planck Institute for Dynamics and Self-Organization (MPIDS), 37077 G\um{o}ttingen, Germany}
\affiliation{Max Planck Institute for the Physics of Complex Systems (MPIPKS), 01187 Dresden, Germany}
\affiliation{Bernstein Center for Computational Neuroscience (BCCN), 37077 G\um{o}ttingen, Germany}
\affiliation{Institute for Theoretical Physics, Technical University of Dresden, 01062 Dresden, Germany}
\affiliation{Department of Physics, Technical University of Darmstadt, 64289 Darmstadt, Germany}

\begin{abstract}
  Across scientific disciplines, thresholded pairwise measures of
  statistical dependence between time series are taken as proxies for the
  interactions between the dynamical units of a network.
  Yet such correlation measures often fail to reflect the underlying physical
  interactions accurately.
  Here we systematically study the problem of reconstructing direct physical
  interaction networks from thresholding correlations.
  We explicate how local common cause and relay structures, heterogeneous
  in-degrees and non-local structural properties of the network generally hinder
  reconstructibility.
  However, in the limit of weak coupling strengths we prove that stationary
  systems with dynamics close to a given operating point transition to universal
  reconstructiblity across all network topologies.
\end{abstract}

\maketitle

\section*{Introduction}
Complex networked systems generate dynamics and thus function that fundamentally
depend on how their units interact \cite{Strogatz2001,Newman2010,Kirst2016}.
As a consequence, knowing the interaction topology of such systems is a key
towards understanding them \cite{Yeung2002, Gardner2003, Yu2006, Timme2007,
  Yu2008, Yu2011, Shandilya2011, Ciofani2012, Nitzan2017}.
Yet, direct access to the topology of physical interactions is largely limited
for many natural systems and across scales, ranging from metabolic and gene
regulatory networks on the subcellular level to neural circuits of millions of
cells, to food webs among organisms and planetary climate networks
\cite{Drossel2002,Shandilya2011,VanBussel2011,Friston2011,Tkacik2011,Rubido2014,Molkenthin2014,Molkenthin2014a,Guez2014,Gilson2016}.
Thus, measures of pairwise statistical dependencies between time series of the
dynamics of their units are often employed as proxies for physical interactions
\cite{Friston2011,Aertsen2012,Timme2014,Nitzan2016,Rubido2014,Opper2015,vanAlbada2015,Gilson2016,Tkacik2011,Wang2016a}.
Assuming sufficiently many and sufficiently accurate data, each such method
provides useful information about how the considered statistical dependency
measures vary across pairs of units.
The value of such a statistical measure, thresholded as desired, e.g. for
significance against coincident correlations, may be taken to quantify the
interactions among these units. 
Yet, such measures themselves do not necessary provide immediate insights into how the
units are directly influencing each other via physical interactions.
In particular, what do correlations generally tell us about direct physical
interactions in network dynamical systems? And is it possible to detect
direct physical interactions among units by thresholding these measures to
reconstruct the topology of the network?

Here, we systematically address this question on a conceptual level and identify
limits of network reconstructibility based on
thresholding pairwise measures of statistical dependence.
In general, non-linearities of intrinsic and coupling dynamics, correlated noise
sources, heterogeneities in time scales and coupling strengths as well as
nontrivial network topology jointly create complex statistical correlation
patterns.
To reveal principal limits of reconstructibility originating from network
interactions (toplogy and strength), we here focus on systems with dynamics
around a given operating point. More specifically, we analyze the idealized setting of
linearly coupled systems with homogeneous dynamical parameters receiving
independent additive noise inputs and evaluate network reconstruction from
thresholding linear correlations obtained from sufficiently long time series.
Reconstruction of physical interactions generally is at
least as hard in any more complex setting, e.g., involving non-linear dynamics
and adequate measures of statistical dependence such as mutual information.
We explicate limits of reconstructibility due to local common cause structures,
local relay structures, topological in-degree heterogeneities as well as
non-local structural elements.
Despite these limitations our analysis interestingly also reveals that,
stationary systems close to operating points exhibit a transition to universal
reconstructibility for sufficiently weak coupling, independent of the
interaction topology.

\section*{Model and Methods}
Consider the dynamics
\begin{equation}
 \tau_\text{gl}\dot{x}_{i}=-x_{i}+\alpha\sum_{j=1}^{N}A_{ij}(x_{j}-x_{i})+\gamma\eta_{i}(t)\label{eq: model}
\end{equation}
of network dynamical systems characterized by variables
$\mathbf{x} = (x_1, \ldots, x_N)$ that interact diffusively with generic
coupling strength $\alpha>0$  on a network topology given by an adjacency matrix
$A$.
The units are driven by independent  white noise $\eta_i(t)$ of strength
$\gamma$ and relax on a time scale $\tau_\text{gl}>0$.
The entries of the weighted adjacency matrix are $A_{ij}>1$ if unit $j$
physically acts on $i$, with all other elements, including the diagonal being
$A_{ij}=0$.
Without loss of generality, we rescale time such that $\tau_\text{gl}=1$.
This dynamics characterizes linear systems as well as stationary systems
sufficiently close to given operating points.

\begin{figure}[!h]
\includegraphics[width=\figwidth]{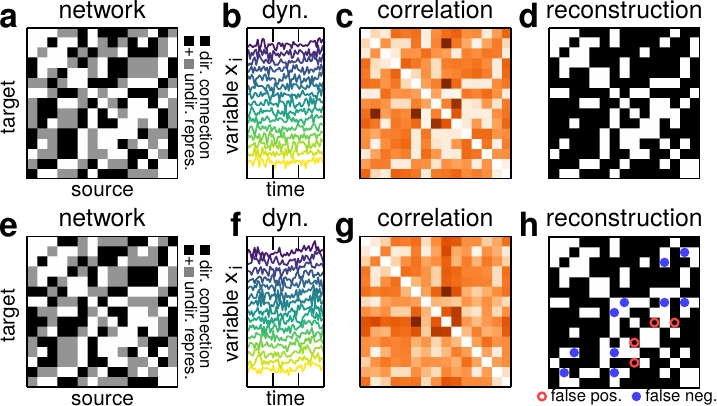}
\caption{
\label{pic: fig1}
(color online)
\textbf{Topology-induced limits of reconstructibility.} 
Reconstructing interaction networks from  correlation thresholding may or may
not yield correct connectivity pattern. (a)-(d) Successful reconstruction of a
network ($N=15$, average degree $\bar{k} = 5$, $\gamma=1$, $\alpha=2, A_{ij} \in \{0,1\}$ for absent and present interactions, resp.).
(e)-(h) Reconstruction of statistically similar network is unsuccessful for any
threshold. (a,e) Adjacency matrix of original network (black indicated directed
interaction, gray undirected network aimed for).  (b,f) Dynamics of the units
yielding (c,g) correlation matrices. Thresholding yields (d) correct or (h)
incorrect reconstruction, depending on the exact topology.}
\end{figure}
 
Can we infer the physical topology from optimally thresholding the matrix
$\CORR$ of pairwise correlations (Fig.~\ref{pic: fig1})? The covariance matrix
$\COV$ defined by the elements
\begin{equation}
  \COV_{ij} = \langle x_ix_j\rangle - \langle x_i\rangle\langle x_j\rangle
\end{equation}
computed using an unbiased time-average $\langle \cdot \rangle$, yields the
correlations
\begin{equation}
 \CORR_{ij} = \frac{\COV_{ij}}{\sqrt{\COV_{ii}\COV_{jj}}}\label{eq: corr}
\end{equation}
by normalization.

Reconstructing the physical topology implies detecting non-zero elements in the
coupling matrix $A$.
Also, as correlation matrices are symmetric by construction, $C_{ij}=C_{ji}$, we
relax the problem to the reconstruction of the undirected representation of the
physical interaction network.
Thus, we aim for the correct reconstruction of the matrix $A'$ the elements of
which are given by

\begin{align}
  \label{eq:undirtopology}
  A'_{ij} =
  \begin{cases}
    1& \textrm{if}\;A_{ij}=1\textrm{ or }A_{ji} = 1\\
    0& \textrm{otherwise}
  \end{cases}
\end{align}
Correlations \eqref{eq: corr} may be thresholded using a (possibly
optimized) threshold $\theta$ to yield an estimate $\hat{A'}$ with elements
$\hat{A'}_{ij}=1$ if $\CORR_{ij}>\theta$ and $\hat{A'}_{ij}=0$ otherwise.
Below we focus on the question whether there is any threshold of the correlation
matrix \eqref{eq: corr} that yields a correct estimate of $A'$.
If there is no such threshold, we call the network non-reconstructible (in this
sense).

The theory of Ornstein-Uhlenbeck processes \cite{Gardiner2009} yields an
analytical expression for the covariance matrix
\begin{equation}
\COV = \gamma^{2}\int_{0}^{\infty}e^{Jt}e^{J^{\textsf{T}}t}\;\dd t . \label{eq: cov_0} 
\end{equation}
Here, the matrix $J$ is given by its elements
\begin{equation}
J_{ij}=\begin{cases}
-(1+\alpha\sum_{j=1}^{N}A_{ij}) & \text{if }i=j\\
\alpha A_{ij} & \text{otherwise}.
\end{cases}\;\label{eq: drift}
\end{equation}
Partial integration of \eqref{eq: cov_0} yields the Lyapunov equation
\begin{equation}
 J\COV+\COV J^\text{T} + \gamma^2I=0\label{eq: lyapunov}
\end{equation}
which we solve numerically \cite{Bartels1972} to obtain the covariance matrix
$\COV$ for arbitrary $(\alpha,\gamma,A)$. Via the relation \eqref{eq: corr}, we
thus semi-analytically obtain all the real-valued elements $C_{ij}$ of the
correlation matrix without any sampling error. We order those to determine
whether there is a threshold $\theta$ separating all existing from all
non-existing links.

\section*{Results}
\subsection*{Topology-induced limits of reconstructibility.}
Even under these idealized conditions, physical interactions are in general not
reconstructible from thresholding the correlation matrix $C$. Whereas some
topologies can be reconstructed via a threshold that separates existing from
absent links (Fig.~\ref{pic: fig1}a-d), many attempted reconstructions yield
false positive and false negative predictions of links, independent of the
threshold (Fig.~\ref{pic: fig1}e-f) and are thus intrinsically
non-reconstructible by correlation thresholding.

\begin{figure}[!h]
 \includegraphics[width=\figwidth]{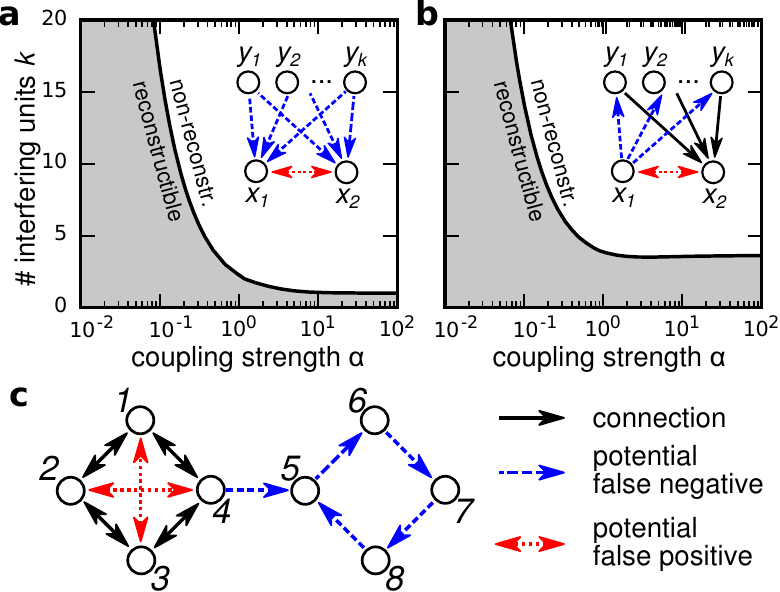}
 \caption{
 \label{pic: fig2}
 (color online)
 \textbf{Topological sources of reconstruction errors and impact of coupling
   strenghts.} 
 (Unspecified parameters as in Fig.~\ref{pic: fig1})
 (a,b) Regions of reconstructible (shaded gray) and non-reconstructible networks
 (white shading) are non-linearly separated for (a) common cause structures and
 (b) relay structures. (Regimes computed by interpolating analytic results using
 \eqref{eq: corr} and \eqref{eq: cov_0}.) (c) Non-local effect renders larger
 networks non-reconstructible: Each circle would be reconstructible alone, but
 the joint network is not ($\alpha=2$).}
\end{figure}

Topologically induced errors and ultimately the limits in reconstructibility can
be of local or of non-local nature (Fig.~\ref{pic: fig2}): For instance, common
input might cause unconnected units to be more correlated than connected units,
a dilemma known as the common cause effect (Fig.~\ref{pic: fig2}a inset).
Likewise, two units may be strongly correlated if the network provides
connectivity between them across a set of intermediate units, thereby forming a
relay structure (Fig.~\ref{pic: fig2}b inset). For both settings,
reconstructibility non-linearly depends on a combination of overall coupling
strength and the number of interfering units in a systematic way
(Fig.~\ref{pic: fig2}a,b, main panels).

In larger networks with diameter $d\geq 3$, additional non-local effects limit
reconstructibility (illustrated in Fig.~\ref{pic: fig2}c). Differences in the
correlation strength may, for instance, be caused by different link densities in
different parts of the network, and imply incorrect link classification.

\begin{figure}[!h]
 \includegraphics[width=\figwidth]{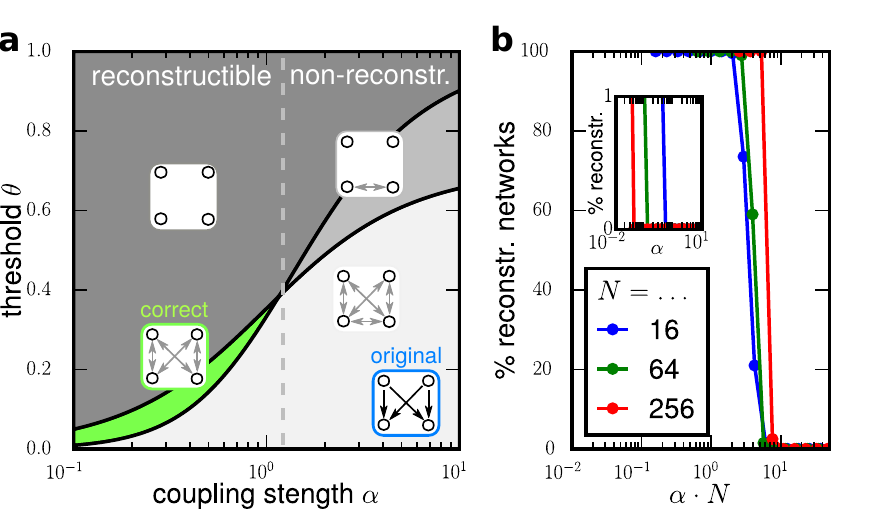}
 \caption{
 \label{pic: fig3}
 \textbf{Transition to reconstructibility for weak coupling.}
 (a) Correlation thresholding yields different estimators (shaded areas with
 graphs as insets) for a given topology (adjacency matrix on bottom right)
 depending on coupling strength and threshold. For sufficiently small coupling
 strength $\alpha$ (left of gray dashed line), there are ideal thresholds
 yielding perfect reconstruction (green shading). (Analytic results obtained
 using \eqref{eq: corr} and \eqref{eq: cov_0}.) (b) Fraction of reconstructible
 networks exhibits transition to full reconstructibility at positive coupling
 strength $\alpha$ (inset) and $\alpha N$ (main panel), illustrated for random
 networks of $N\in \{16, 64, 256\}$ units and link probability $p=0.5$. Every
 arbitrary network exhibits such a transition individually (see text).}
\end{figure}

\subsection*{Universal transition to non-reconstructibility.}
The coupling strength $\alpha$ controls the impact of both, local and non-local
influences on reconstructibility. For instance, analytic treatment of a small
common cause structure (Fig.~\ref{pic: fig3}) reveals that the system becomes
reconstructible for all sufficiently small coupling strengths $\alpha$  while it
is non-reconstructibility if $\alpha$ is too large. This systematic transition
prevails for any number of common input units in common cause structures as well
as for any number of relay units in relay structures (See Supplementary material
for detailed derivations).

Interestingly, all topology-induced limits disappear for sufficiently weak
coupling, as seen from the following analytic argument: Rewriting the matrix
\begin{equation}
  \label{eqn:JfromL}
  J=-(\unity + \alpha L)
\end{equation}
in terms of the graph Laplacian $L$ with elements
\begin{equation}
  \label{eqn:L}
  L_{ij}=-A_{ij} + \delta_{ij}\sum_j A_{ij}
\end{equation}
(where $\delta_{ij}=1$ if $i=j$ and zero otherwise is the Kronecker-delta) and
expanding \eqref{eq: cov_0} for $\alpha\ll1$ yields
\begin{align}
\begin{split}
 \COV=\frac{\gamma^2}{2}&\left[\unity -\frac{\alpha}{2}(L+L\T)\right.\\
 &+\left.\frac{\alpha^2}{2}\left( LL\T+\frac{L^2+L\T^2}{2} \right) \right]+\orderof{\alpha^3}\;.
\end{split}\label{eq: cov extension}
\end{align}
The term $\alpha(L+L\T)/2$ on the r.h.s.~of \eqref{eq: cov extension} does only
contribute to entries $\sigma_{ij}$ that reflect existing links because
otherwise $L_{ij}=A_{ij}=0$. Thus, the covariance of coupled units scales
linearly with $\alpha$ whereas for uncoupled units it scales quadratically. So
for sufficiently small coupling strength $\alpha$, covariances of coupled units
will be larger than those of uncoupled units. This result transfers to the
elements of the correlation matrix $\CORR$ in \eqref{eq: corr} because diagonal
elements of the covariance matrix $\COV$ are of order
\begin{equation}
\label{eqn:sigmadiag}
\sigma_{ii}=\orderof{\alpha^0} \textrm{ as } \alpha \rightarrow 0.
\end{equation}
Hence, every network topology is reconstructible for sufficiently small coupling
strengths.

\subsection*{Illustrative example of reconstructibility transition.}
Furthermore, specific families of networks with homogeneous connectivity are
reconstructible via correlation thresholding for all coupling strengths, weak
and strong. 

As we demonstrate for illustration, this is the case for directed ring like topologies with
$\bar{k}$ neighbors.
In these networks the correlation matrix $\CORR$ is strictly proportional to the
covariance matrix $\COV$ so that it is sufficient to show reconstructibility
with respect to the covariance matrix.
Also, since the covariance matrix $\COV$ is a circulant, it is sufficient to
show reconstructibility only for the connections of one unit.
The reconstructibility conditions is identical for all units.
For simplicity of presentation, we take the number $N$ of units to be even.

We order the units in such a way that it reflects the network topology, i.e.
\begin{equation}
  A_{i, (i+l) \Mod N} = \begin{cases}
    1 & \textrm{if}\;1<l\leq \bar{k}\\
    0 & \textrm{otherwise}
  \end{cases}\;,
\end{equation}
and replace $J=-(\unity+\alpha A)$ in Eq.~\eqref{eq: lyapunov} to obtain
\begin{align}
  \label{eq:cirlcelyapunov}
  \sum\limits_{l=1}^{\bar{k}} \COV_{i,i+n-l} - 2\frac{\alpha+{\bar{k}}}{\alpha}\COV_{i,i+n}+ \sum\limits_{l=1}^{\bar{k}} \COV_{i,i+n+l} = -\frac{\gamma^2}{\alpha}\delta_{i,i+n}\;
\end{align}
for the covariance matrix $\COV$. Here, the index $i$ indicates the number of
the unit and is thus arbitrary.

Transforming this equation into Fourier space yields
\begin{equation}
 \sum_{l=1}^{k} e^{-2\pi i \frac{lm}{N}} s_m - 2(\frac{1}{\alpha}+k)s_m + \sum_{l=1}^{k} e^{2\pi i \frac{lm}{N}} s_m = -\frac{\gamma^2}{\alpha}
\end{equation}
with solution 
\begin{equation}
  s_m = \frac{\gamma^2}{\alpha}\frac{1}{2(\frac{1}{\alpha}+k) -2\sum\limits_{l=1}^k \cos\left(2\pi \frac{lm}{N}\right)}
\end{equation}
in Fourier coordinates.
An inverse Fourier transformation yields the analytic solution

\begin{align}
  \label{eq:cirlceinversefourier}
  \COV_{i,i+n}&= \frac{\gamma^2}{2+2\alpha \bar{k} + \alpha}\left\{\delta_{0n}+\sum_{l=1}^\infty \frac{\alpha^l\zeta_{{\bar{k}},n}^{\ast l}}{(2+2\alpha \bar{k} + \alpha)^l}\right\}
\end{align}
where the sequences $\zeta_{{\bar{k}},n}^{\ast l}$ are repeated convolutions of
the step sequence
\begin{align}
  \zeta_{\bar{k},n} &=
  \begin{cases}
   1 & \text{if}\quad n\!\!\!\mod\!N \leq \bar{k}\\
   1 & \text{if}\quad N-\bar{k}\leq n\!\!\!\mod\!N\\
   0 & \text{otherwise}
 \end{cases}\;,
\end{align}
i.e.,
\begin{align}
\begin{aligned}
 \zeta_{\bar{k}}^{\ast l} :&= (\zeta_{\bar{k}} \ast\zeta_{\bar{k}}^{\ast (l-1)})\;, &\zeta_{\bar{k}}^{\ast1} &= \zeta_{\bar{k}}\;.
\end{aligned}
\end{align}

Since the sequences $\zeta_{\bar{k},n}^\ast$ are monotonically decreasing in the
interval $n\in[-N/2,N/2]$ covariance only decreases with distance in the
circular graph.
Because for any given unit $i$, connected units are closer than non-connected
units, for every such network with $k$-regular topology, a threshold exists that
separates existing from absent links, making these networks reconstructible for
arbitrary coupling strengths, for any network size $N$ and for any number of
neighbors $\bar{k}<\frac{N}{2}$.
For $\bar{k}=\frac{N}{2}$ the undirected representation of the network is fully
connected and reconstruction is trivial.

\begin{figure}[!h]
 \includegraphics[width=\figwidth]{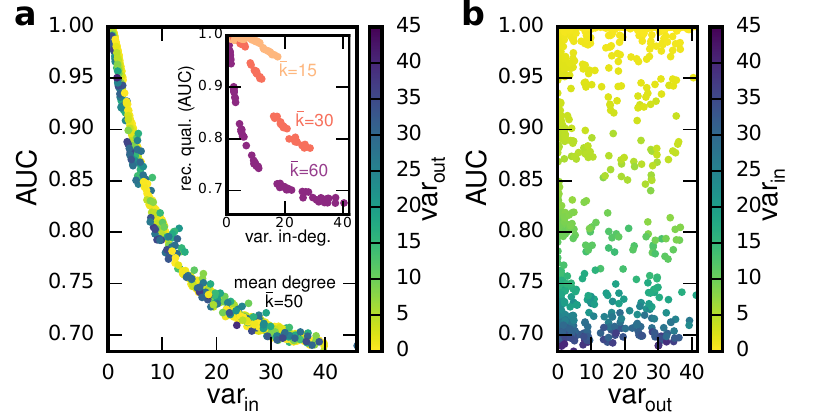}
  \caption{
  \label{pic: fig4}
  (color online)
  \textbf{Reconstruction systematically varies with heterogeneities in
    in-degree, but not in out-degree.} (a) $\AUC$ exhibits functional dependency
  on the variance of the in-degree distribution $\textrm{var}_\textrm{in}$,
  regardless of the variance of the out-degree $\textrm{var}_\textrm{out}$.
  Inset: Qualitative behavior is the same for differnet mean degrees. (b) No
  significant dependency of reconstruction quality on out-degree heterogeneity
  (network size $N=150$ throughout, $\alpha=1$, $A_{ij}\in\{0,1\}$).}
\end{figure}

\subsection*{Which heterogeneities hinder reconstruction?}
Given the insights from the ring-like networks, we hypothesized that if
topological irregularities increase, they decrease and ultimately hinder network
reconstructibility.
To analyze the overall impact of topology on reconstruction quality, we
investigated ensembles of directed networks in the regime between regular and
random, employing a modified Watts-Strogatz small world model \cite{Watts1998}:
Starting with a regular ring of $N$ units with each unit receiving directed
links from $\bar{k}$ preceding nodes, the source and the target of each link are
detached with probability $q_\text{out}$ and probability $q_\text{in}$
respectively. The resulting loose ends are randomly redistributed in the network
while avoiding self-loops and multiple links. This creates networks of mean
degree $\bar{k}$ whose in-degree distribution $p^\text{in}_k$ and out-degree
distribution $p^\text{out}_k$ are altered separately from their original values
$p^\text{in}_k=p_k^\text{out}=\delta_{k\bar{k}}$ by varying $q_\text{in}$ and
$q_\text{out}$. This random graph ensemble contains networks with unimodal
degree distributions (binomial for $q_\text{in}=q_\text{out}=1$, $\bar{k}\ll N$
and $1\ll N$) so that the variances of the distributions serve as indicators for
the inhomogeneities in the network.

Considering a fixed coupling strength (e.g., $\alpha=1$), we quantify
reconstructibility by measuring the AUC, the area under the ROC (receiver
operating characteristic) curve, generated by a variable correlation threshold
$\theta$. AUC ranges from AUC=0.5 for random guessing to AUC=1 for perfect
reconstructibility (see Supplemental Information for an introduction to ROC
curves). For networks that are not densely connected ($\bar{k}<(N-1)/2$), we
find that reconstruction quality systematically decreases with in-degree
heterogeneity, with the AUC exhibiting a functional dependency on the variance
of the in-degree distribution, yet is almost independent of the variance of the
out-degree distribution (compare Fig.~\ref{pic: fig4}a with Fig.~\ref{pic:
  fig2}b). Thus, the reconstruction error is mainly explained by the in-degree
heterogeneity. We obtain qualitatively similar results across different average
connectivities $\bar{k}$ (inset of Fig.~\ref{pic: fig4}a).

\section*{Conclusions}
In summary, we have systematically investigated reconstructibility of physical
interaction networks from thresholding statistical correlations.
Beyond valuable previous studies which targeted the impact of correlated noise
and estimation errors \cite{Bialonski2011, Bialonski2013}, we revealed intrinsic
limits of reconstructibility induced by the strengths of network interactions
and their topology.
In particular, a number of distinct topological factors contribute in a
systematic way: local common cause structures, local relay structures, in-degree
heterogeneities as well as non-local structural elements of a network resulting
from different link densities in different network parts.
Intriguingly, for stationary dynamics and arbitrary network topologies we
uncovered a transition to full reconstructibility when decreasing the coupling
strengths.
Whereas the exact critical coupling strength to transition to reconstructibility
depends on the topology, it is guaranteed to occur for all
topologies.

Given the limitations of correlation thresholding, alternate methods of
reconstruction from time series data are required.
For systems that are strongly non-linear and non-stationary, the range of
inference methods is currently largely limited to systems with models known a
priori.
Such non-linear systems in general pose a number of additional challenges,
including that there typically is no well-defined, temporally constant coupling
strength between the units.
Future studies would need to investigate model-independent methods to obtain
physical interaction structure from recorded non-linear dynamics
\cite{Yeung2002, Gardner2003, Yu2006, Timme2007, Yu2008, Yu2011, Shandilya2011,
  Ciofani2012, Nitzan2017}.
Our main result on full reconstrucatbility in the weak coupling limit might
provide a useful initial step towards the reconstruction of non-linear and
non-stationary networks: By systematically combining localized but faithful
reconstructions obtained from an entire set of dynamics around different
operation points in weakly coupled networks a global picture of the underlying
interactions and their network state-dependencies could be obtained.

Our results on topology-induced limits of network reconstructibilty not only
further our theoretical insights about the relations between statistical
correlation and physical interaction networks \cite{Timme2014, Nitzan2016, Witt} but also
indicate where principal care has to be taken in applications when analyzing
statistical correlation data to reveal aspects of direct physical interactions.

\section*{Acknowledgments}
We thank E. Ching, A.-L. Barabasi and G. Yan for valuable discussions. This work
was supported by the Max Planck Society (BL, MT), the Germany Ministry for
Education and Research (BMBF) under grant no. 01GQ1005B (CK, MT), and an
independent research fellowship by the Rockefeller University, New York, USA
(CK).
Supported through the German Science Foundation (DFG) by a grant towards the
Center of Excellence `Center for Advancing Electronics Dresden' (cfaed).

\nolinenumbers

\onecolumngrid
\vspace{\fill}

\appendix*


\section{Model}
\label{sec:model}

Here, we consider networks of $N$ units each described by a state variable
$x_i$, $i\in\{1,\ldots,N\}$, that evolve according to an Ornstein-Uhlenbeck (OU)
process given by
\begin{align}
  \label{eq:model}
  \dot{x_i}=-x_i+\alpha \sum\limits_{j=1}^N A_{ij} (x_j-x_i) + \gamma\eta_i(t)\;
\end{align}
with $\boldsymbol{\dot{x}}, \boldsymbol{x}\in\mathbb{R}^N$, white noise vector
$\boldsymbol{\eta}(t)\in\mathbb{R}^N$, adjacency matrix
$A_{ij}\in\{0,1\}^{N\times N}$, coupling strength $\alpha\in\mathbb{R_+}$ and
noise strength $\gamma\in\mathbb{R}_+$.

Introducing the Laplace matrix $L$ with elements
\begin{align}
  L_{ij} = -A_{ij} +\delta_{ij}\sum\limits_{k=1}^NA_{ik}
\end{align}
(where $\delta_{ij}$ is the Kronecker-delta) and the drift matrix
\begin{align}
  J = -(\unity + \alpha L) 
\end{align}
the process \eqref{eq:model} can be rewritten in the multivariate form
\begin{align}
  \dot{\boldsymbol{x}}=J\boldsymbol{x}+\gamma\boldsymbol{\eta}(t).
\end{align}

Since the drift matrix $J$ is diagonally negative dominant, it has only
eigenvalues with non-zero negative real part, so that the process has a stationary solution with covariance matrix 
\begin{align}
\label{eq:integral}
 \COV &= \gamma^2\int\limits_0^\infty e^{Jt}e^{J\T t}\;\dd t
\end{align}
that fulfills the Lyapunov equation
\begin{align}
  \label{eq:lyapunov}
 J\COV + \COV J\T + \gamma^2\unity = 0\;.
\end{align}
For reference see \cite{Gardiner2009}.

The existence of an analytic equation for the covariance matrix $\COV$ enables
us to compute the covariance matrix directly without simulating the
process, avoiding additional errors induced by finite time series.

\section{Detailed Analytic Derivation of Correlations}
Here, we present the detailed analytic derivation of the analytic correlations in the generalized
common cause problem and the generalized relay structure problem.

We proceed as follows:\\
First, we compute the instantaneous covariance matrix $\COV$ of the OU process
by solving the integral given by \eqref{eq:integral}, or more precisely
\begin{align}
  \COV = \frac{\gamma^2}{\alpha}\int\limits_0^\infty e^{-\frac{2}{\alpha} t'} \underbrace{e^{-Lt'}e^{-L\T t'}}_{=:\Lambda(t')} \;\dd t'\;.
\end{align}
For this purpose, we calculate the matrix $\Lambda(t)$, which is determined by the
topology, and integrate element-wise to get elements of the matrix $\COV$.\\
Then, we compute the Pearson correlation matrix $\CORR$ using its definition \mbox{$\CORR_{ij} = \frac{\COV_{ij}}{\sqrt{\COV_{ii}\COV_{jj}}}$}.

For Fig. 2a,b in the manuscript, we then calculate the difference in correlation for existing
connections and non-existing connections as a function of coupling
strength $\alpha$ and number of source units (common cause problem) $m$ or
transmitting units (relay structure) $m$ and interpolate the zero-crossing of
this difference in $\alpha$-$m$ space
numerically.

\subsection{Common Cause Structure}

Let $\boldsymbol{Y}=(Y_1,Y_2,\ldots,Y_m)\in\mathbb{R}^m$,
$\boldsymbol{X}=(X_1,X_2)\in\mathbb{R}^2$ be two vectors of unit representing
random variables and let each element of $\boldsymbol{Y}$ be a source unit of
each element of $\boldsymbol{X}$.
Then, the topology $A$ and the Laplacian $L$ for the network of the process
$\boldsymbol{Z}=(\boldsymbol{X},\boldsymbol{Y)}$  are given by
\begin{align}
A = 
\begin{pmatrix}
0 &0 &1 &1 &\cdots\\
0 &0 &1 &1 &\cdots\\
0 &0 &0 &0 &\cdots\\
\vdots &\vdots &\vdots &\vdots &\ddots
\end{pmatrix}
\quad\Rightarrow\quad
L = 
\begin{pmatrix}
m &0 &-1 &-1 &\cdots\\
0 &m &-1 &-1 &\cdots\\
0 &0 &0  &0 &\cdots\\
\vdots &\vdots &\vdots &\vdots &\ddots
\end{pmatrix}\;.
\end{align}
The matrix power of $L$ yields
\begin{align}
 L^n = \begin{cases}
	  m^{n-1} L & n\neq0\\
	  \unity    & n=0
       \end{cases}\quad n\in\mathbb{N}\;.
\end{align}
Thus, the matrix exponential is given by
\begin{align}
e^{-Lt} &= \sum_{n=0}^\infty \frac{(-t)^n}{n!}L^n\nonumber\\
        &= \unity + \sum\limits_{n=1}^\infty \frac{(-t)^nm^{n-1}}{n!} L \nonumber\\
        &=  \unity + \frac{e^{-mt}-1}{m}L\;.
\end{align}
Hence,
\begin{align}
\Lambda(t):=e^{-Lt} e^{-L\T t} &= \unity + \frac{e^{-mt}-1}{m}(L+L\T) + \left(\frac{e^{-mt}-1}{m}\right)^2LL\T\;
\end{align}
with
\begin{align}
 LL^T = 
\begin{pmatrix}
m^2+m &m &0 &\cdots\\
m &m^2+m &0 &\cdots\\
0 &0 &0  &\cdots\\
\vdots &\vdots &\vdots &\ddots
\end{pmatrix}\;,
\end{align}
so that the entries of $\Lambda$ are given by
\begin{align}
\Lambda_{11} &= \Lambda_{22} = 1 + 2(e^{-mt}-1)+\frac{m^2+m}{m^2}(e^{-mt}-1)^2\\
\Lambda_{33} &= \ldots = \Lambda_{NN} = 1\\
\Lambda_{12} & = \frac{(e^{-mt}-1)^2}{m}\\
\Lambda_{13} &= \ldots = \Lambda_{1N} = \Lambda_{23} = \ldots = \Lambda_{2N} = \Lambda_{13} = -\frac{e^{-mt}-1}{m}\;.
\end{align}
All remaining entries  not defined by $\Lambda = \Lambda\T$ are zero.\\
Integrating 
\begin{align}
 \COV_{ij} = \frac{\gamma^2}{\alpha}\int\limits_0^\infty e^{-\frac{2}{\alpha} t}\Lambda_{ij}(t) \;\dd t
\end{align}
yields
\begin{align}
  \COV_{11} &= \COV_{22} = \gamma^2 \frac{\alpha^2 m+\alpha m + 2}{(\alpha 2 + 2)(2\alpha m +2)}\\
 \COV_{33} &= \ldots = \COV_{NN} = \frac{\gamma^2}{2}\\
  \COV_{12} &= \gamma^2 \frac{\alpha^2 m}{(\alpha m +2)(2\alpha m +2)}\\
 \COV_{13} &= \ldots = \COV_{1N} = \COV_{23} = \ldots = \COV_{2N} = \COV_{13} = \frac{\gamma^2}{2}\frac{\alpha m }{\alpha m + 2}\;.
\end{align}
Normalizing yields two different correlation values: The correlation 
\begin{align}
  \CORR_{xx} &= \frac{\alpha^ 2 m}{\alpha^2m+\alpha m +2}    \label{eq: common cause false positive}
\end{align}
of the non-connected nodes $X_1$ and $X_2$ and the correlation
\begin{align}
  \CORR_{xy} &= \sqrt{\left(\frac{\alpha m +2}{\alpha m +4} \right)\left( \frac{\alpha^2}{\alpha^2m +\alpha m + 2} \right)}\label{eq: common cause false negative}
\end{align}
for connection from units in $\boldsymbol{Y}$ to units in $\boldsymbol{X}$.

For Fig. 2a of the main article, we determined the difference
between correlations of unconnected pairs and connected pairs  $C_{xx}-C_{xy}$
in dependence on the coupling strength $\alpha$ and the number of source units
$m$ and plotted the zero crossing in $\alpha$-$m$ space. This curve marks the
transition from reconstructible to non-reconstructible.

\subsection{Relay Structures}
We perform the same analysis that was done for the common cause structure (see above) for the relay structure. 
\\\\
Here, we define $\boldsymbol{Z}=(X_2,\boldsymbol{Y},X_1)\T$. Each element of $\boldsymbol{Y}$ gets inputs from $X_1$ and each element of $\boldsymbol{Y}$ is a source unit of $X_2$.\\
The adjacency matrix and the Laplacian of the network for $\boldsymbol{Z}$ are
\begin{align}
A = \underbrace{
\begin{pmatrix}
0 &1 &\cdots &1 &0\\
0 &0 &\cdots &0 &1\\
\vdots &\vdots &\ddots &\vdots &\vdots\\
0 &0 &\cdots &0 &1\\
0 &0 &\cdots &0 &0
\end{pmatrix}}_{m+2}
\qquad\Rightarrow\qquad
L = 
\begin{pmatrix}
m &-1 &\cdots &-1 &0\\
0 &1 &\cdots &0 &-1\\
\vdots &\vdots &\ddots &\vdots &\vdots\\
0 &0 &\cdots &1 &-1\\
0 &0 &\cdots &0 &0
\end{pmatrix}\;.
\end{align}
The matrix power of the Laplacian yields
\begin{align}
L^n = 
\begin{pmatrix}
m^n &-\frac{1-m^n}{1-m} &\cdots &-\frac{1-m^n}{1-m} & \frac{m-m^n}{1-m}\\
0 &1 &\cdots &0 &-1\\
\vdots &\vdots &\ddots &\vdots &\vdots\\
0 &0 &\cdots &1 &-1\\
0 &0 &\cdots &0 &0
\end{pmatrix}\;,
\end{align}
where used the geometric series.\\
Hence, the matrix exponential is given by
\begin{align}
e^{-Lt} =  
\begin{pmatrix}
e^{-mt} & \frac{e^{-t}-e^{-mt}}{m-1} &\cdots &\frac{e^{-t}-e^{-mt}}{m-1} & \frac{m(1-e^{-t})-1+e^{-mt}}{m-1}\\
0 &e^{-t} &\cdots &0 &1-e^{-t}\\
\vdots &\vdots &\ddots &\vdots &\vdots\\
0 &0 &\cdots &e^{-t} &1-e^{-t}\\
0 &0 &\cdots &0 &1
\end{pmatrix}\;. 
\end{align}
The matrix $\Lambda$ and the covariance matrix $\COV$ are computed following the
same ideas as in the previous paragraph.\\
We find four correlation values: Two for the existing connections $X_1\rightarrow Y_i$
\begin{align}
  \CORR_{xy} &= \sqrt{\left( \frac{1+\alpha}{2+\alpha}\right)\left(\frac{\alpha^2}{\alpha^2+\alpha+2}\right)}\label{eq: relay structure false negative} 
\end{align}
and $Y_i\rightarrow X_1$
 \begin{alignat}{1}
 \begin{aligned}
   \CORR_{yx} &=
\alpha \left(\alpha^3(m^2+m)+4\alpha^2 m+2\alpha (m+1) + 4\right)\sqrt{\alpha m +1}\cdot\\
&\sqrt{(\alpha m +2)(\alpha^2+\alpha+2)(\alpha m+\alpha + 2)(\alpha^5(m^3+m^2)+}\\
&\overline{\alpha^4 m(5m+1)+\alpha^3(5m^2+9m+2)+2\alpha^2(m^2+9m+5)+8\alpha(m+2)+8)}^{-1}
   \label{eq: relay structure}
 \end{aligned}
 \end{alignat}
 and two for the non-existing connections $Y_i\leftrightarrow Y_j$
 \begin{align}
  \CORR_{yy} &= \frac{\alpha^2}{\alpha^2+\alpha+2}
 \end{align}
 and $X_1\leftrightarrow X_2$
 \begin{alignat}{1}
   \begin{aligned}
     \CORR_{xx} &= \alpha ^2m\sqrt{(  \alpha  + 1  )( \alpha m + 1 )( \alpha m + \alpha  + 2 )( \alpha 
       + 2 )( \alpha m + 2 )}\sqrt{ \alpha ^5m^2( m+ 1 ) + }\\
     &\overline{\alpha ^4m(5m + 1) + \alpha ^3(5m^2 + 9m + 2) + 2\alpha ^2(m^2 + 9m + 5) + 8\alpha (m + 2) + 8 }^{-1}
   \end{aligned}
 \end{alignat}

As for common cause structures, we compute the difference between 
the correlation of unconnected units $C_{xx}$ and the smallest correlation
among connected units $C_{xy}$ and determine the zero-crossing in $\alpha$-$m$
space. Like before, this curve marks the transition from reconstructible to non-reconstructible.

\section{Reconstructibility in the Weak Coupling Limit}
Resolving $J=-(\unity+\alpha L)$ in \eqref{eq:integral} yields
\begin{align}
 \COV &= \gamma^2 \int\limits_0^\infty e^{-2t} e^{-\alpha L t}e^{-\alpha L\T t}\;\dd t\;.
\end{align}
Since the matrix exponential is defined as 
\begin{align}
 e^{-\alpha L t} &:= \sum_{n=0}^\infty \frac{\alpha^n t^n}{n!} L^n\\
 &= \unity + \alpha t L + \frac{\alpha^2 t^2}{2}L^2 + \orderof{\alpha^3}\;
\end{align}
with finite rest $\orderof{\alpha^3}$, the integral can be written as
\begin{align}
\begin{split}
\COV &=\gamma^2\int\limits_0^\infty
\exp(-2t)
\left(\unity-\alpha L t+\frac{\alpha^2t^2}{2}L^2+\ldots\right)\\
&\quad\left(\unity-\alpha L\T t+\frac{\alpha^2t^2}{2}L\T^2+\ldots\right)\;\dd t
\end{split}\\
\begin{split}
 &=\gamma^2\int\limits_0^\infty
\exp(-2t)
\bigg(
\unity
-\alpha (L+L\T) t\\
&\quad+\frac{\alpha^2t^2}{2}(2LL\T+L^2+L\T^2)\bigg)
+\orderof{\alpha^3}
\;\dd t
\end{split}\\
\begin{split}
&=\gamma^2\bigg\{\frac{1}{2}\unity -\frac{\alpha}{4}(L+L\T)\\
&\qquad+\frac{\alpha^2}{8}(2LL\T + L^2+L\T\,^2) +\orderof{\alpha^3} \bigg\} 
\end{split}\;.
\end{align}
Hence, diagonal elements of the covariance matrix $\COV$  are given by
\begin{align}
 \COV_{ii} = \frac{\gamma^2}{2}+\orderof{\alpha^1}\;,
\end{align}
elements corresponding to links are given by 
\begin{align}
 \COV_{ij}^\text{c}= -\frac{\gamma^2\alpha}{4}(L_{ij}+L_{ji})+\orderof{\alpha^2}\;,
\end{align}
and elements corresponding to non-links are given by
\begin{align}
 \COV_{kl}^\text{nc}= \frac{\gamma^2\alpha^2}{8}\overbrace{(2LL\T + L^2+L\T\,^2)_{kl}}^{M_{kl}}+\orderof{\alpha^3}\;.
\end{align}
Hence, elements of the correlation matrix $\CORR$ belonging to connections are given by
\begin{align}
 \CORR_{ij}^\text{c} = -\frac{1}{2}\frac{\alpha(L_{ij}+L_{ji})+\orderof{\alpha^2}}{1+\orderof{\alpha^1}}
\end{align}
and elements of the correlation matrix $\CORR$ corresponding to non-connections are given by
\begin{align}
 \CORR_{kl}^\text{nc} = \frac{1}{4}\frac{\alpha^2M_{kl}+\orderof{\alpha^3}}{1+\orderof{\alpha^1}}\;.
\end{align}
For weak coupling strength $\alpha\ll1$ this ensures that there is a critical
coupling strength $\alpha_c(A)$ for which every coupling strength
$\alpha<\alpha_c(A)$ results in $\CORR_{ij}^\text{c}>\CORR_{kl}^\text{nc}$ for
all indices $i,j,k,l$.
Hence, there exists a threshold $\theta(\alpha,A)$ for the correlation matrix
$\CORR$ that results in the reconstruction of the original network $A$.

\section{Reconstructibility of Circles}
We proof that any directed circular topology results in
a correlation matrix $\CORR$ that can be thresholded such that the original
network topology $A$ is retrieved.
Hence, any circular topology is reconstructible by correlation thresholding.

The proof goes as follows:
\begin{enumerate}
\item We demonstrate that the correlation between units decreases monotonically
  with distance in the circle. 
\item We show that every unit is more correlated with its farthermost connected
  unit than with its closest unconnected unit.
\item We conclude that every pair of connected units is stronger correlated than
  any pair of non-connected units such that the network is reconstructible by
  correlation thresholding.
\end{enumerate}

\subsection{Proof of Monotonicity}
\label{sec:proof of monotonicity}

From \eqref{eq:lyapunov} we obtain 
\begin{align}
  \COV_{ij} &= \frac{1}{2+\alpha(k_{\textrm{in},i}+k_{\textrm{in},j})}\left(\gamma^2\delta_{ij} + \alpha\left[\sum\limits_{\{l:i\leftarrow l\}} \COV_{jl} + \sum\limits_{\{l:j\leftarrow l\}}\COV_{li}\right]\right)\;,
\end{align}
as a relation between elements of the covariance matrix $\COV$.
Here, $\delta_{ij}$ is the Kronecker-delta, $k_{\textrm{in},i}$ ist the
in-degree of unit $i$ and $\sum\limits_{\{l:i\leftarrow l\}}$ is the sum over
all indices of units that are in-neighbors of unit $i$.

The topology of the network determines how to resolve the two sums. In case of
directed $k$-rings each units gets input from the subsequent $k$ units. In
addition, the in-degree for each node is $k$. Hence,
\begin{align}
  \label{eq:circular self equation}
  \COV_{ij} &= \frac{1}{2+2\alpha k}\left(\gamma^2\delta_{ij} + \alpha\left[\sum\limits_{l=1}^k \COV_{j,i+l} + \sum\limits_{l=1}^k\COV_{j+l,i}\right]\right)\;.
\end{align}
In a $k$-ring $k$ is the maximum distance between connected units, for this
reason $2k+1<N$. Equality denotes a network in which all units are already
connected either by incomming or outgoing connections, so that a reconstruction
is trivial because no unconnected pairs exist.

The topological features of a $k$-ring have further consequences:
Due to the fact that such a graph is rotationally invariant, the covariance between two
units only depends on the distance in the ring. Thus, $\COV$ is a circulant
matrix, i.e. $\COV_{(i+n)\mod N,(j+n)\mod N} = \COV_{ij}$ for all
$n\in\mathbb{Z}$. This means, $\COV$ is fully determined by the sequence
$(\COV_{i,i+n})_{n=0}^{N-1}$. Also, the correlation values
$\CORR_{ij}:=\frac{\COV_{ij}}{\sqrt{\COV_{ii}\COV{jj}}}=\frac{\COV_{i,i+n}}{\COV_{ii}}$
are just proportional to the covariance values. Hence, \emph{thresholding covariance
is fully equivalent to thresholding correlation.}

For convenience, we define the periodic sequence
$\varkappa\hat{=}(\varkappa_n)_{n=-\infty}^{\infty}$ with period $N$ and $\varkappa_n:=\COV_{i,i+n}$.
This sequence fulfills $\varkappa_{n+N} = \varkappa_n$ due to periodic boundary
conditions for the indices.  
In addition, the covariance matrix $\COV$ is symmetric, i.e. $\COV_{ij}=\COV_{ji}$, so that the
periodic sequence $\varkappa$ also has to fulfill $\varkappa_n=\varkappa_{-n}$
for all $n\in\mathbb{Z}$.

Using both symmetries \eqref{eq:circular self equation} yields
\begin{align}
  \sum\limits_{l=1}^k \COV_{i,i+n-l} - 2(\frac{1}{\alpha}+k)\COV_{i,i+n}+ \sum\limits_{l=1}^k \COV_{i,i+n+l} &= -\frac{\gamma^2}{\alpha}\delta_{i,i+n}\\
  \Rightarrow \sum\limits_{l=1}^k\varkappa_{n-l}-2(\frac{1}{\alpha}+k)\varkappa_n + \sum\limits_{l=1}^k\varkappa_{n+l} &= -\frac{\gamma}{\alpha}\delta_{0m} \label{eq:sequencesum}
\end{align}

We make use of the periodicity of $\varkappa$ by applying the Fourier transform $s:=\mathcal{F}[\,\varkappa\,]$.\\
Multiplying \eqref{eq:sequencesum} by $e^{-2\pi i\frac{nm}{N}}$ and summing the resulting equation over all \mbox{$m\in[0,N-1]$} yields
\begin{align}
 \sum\limits_{m=0}^{N-1}\left\{\sum\limits_{l=1}^k \varkappa_{n-l}e^{-2\pi i \frac{nm}{N}} - 2(\frac{1}{\alpha}+k) \varkappa_{n}e^{-2\pi i \frac{nm}{N}}  + \sum\limits_{l=1}^k \varkappa_{n+l}e^{-2\pi i \frac{nm}{N}}\right\} &= -\frac{\gamma^2}{\alpha}\\
 \Rightarrow\quad \sum\limits_{l=1}^k\left\{ \sum_{m=0}^{N-1} \varkappa_{n-l} e^{-2\pi i \frac{nm}{N}} - \frac{2(\frac{1}{\alpha}+k)}{k} \sum_{m=0}^{N-1} \varkappa_n e^{-2\pi i \frac{nm}{M}} + \sum_{m=0}^{N-1} \varkappa_{n+l} e^{-2\pi i \frac{nm}{N}}\right\} &= -\frac{\gamma^2}{\alpha}\\
 \Rightarrow\quad \sum_{l=1}^{k} e^{-2\pi i \frac{lm}{N}} s_m - 2(\frac{1}{\alpha}+k)s_m + \sum_{l=1}^{k} e^{2\pi i \frac{lm}{N}} s_m &= -\frac{\gamma^2}{\alpha}\\
 \Rightarrow\quad s_m = \frac{\gamma^2}{\alpha}\frac{1}{2(\frac{1}{\alpha}+k) -2\sum\limits_{l=1}^k \cos\left(2\pi \frac{lm}{N}\right)}\;.
\end{align}

\subsubsection{Inverse Fourier Transform $\varkappa = \mathcal{F}^{-1}[\,s\,]$}\label{apdx: ring fourier her}

We rewrite $s_m$ to get
\begin{align}
 s_m &= \mathcal{F}[\,\varkappa\,]_m\nonumber\\
     &= \frac{\gamma^2}{\alpha}\frac{1}{(\frac{2}{\alpha}+2k+1)- \underbrace{\left(2\sum\limits_{l=1}^k \cos\left(2\pi \frac{lm}{N}\right) + 1\right)}_{:=z_{k,m}} }\nonumber\\
     &= \frac{\gamma^2}{\alpha(\frac{2}{\alpha}+2k+1)}\left(1-\frac{z_{k,m}}{\frac{2}{\alpha}+2k+1}\right)^{-1}\nonumber\\
     &= \frac{\gamma^2}{\alpha(\frac{2}{\alpha}+2k+1)}\sum\limits_{l=0}^\infty\left(\frac{z_{k,m}}{\frac{2}{\alpha}+2k+1}\right)^l\;.
\end{align}
Here, we used the geometric series and the fact that $ |z_{k,m}| < \frac{2}{\alpha}+2k+1$ for all $\alpha<\infty$.\\
$z_k\widehat{=}\left(z_{k,m}\right)_{m=-\infty}^\infty$ is a periodic sequence the inverse Fourier transform of which $\zeta_k:=\mathcal{F}^{-1}[\,z_k\,]$ yields
\begin{align}
 \zeta_{k,n} &=  \mathcal{F}^{-1}[\,z_k\,]_n\nonumber\\
 &= \frac{1}{N} \sum_{m=0}^{N-1} z_{k,m} e^{2\pi i \frac{nm}{N}}\nonumber\\
 &= \frac{1}{N} \sum_{m=0}^{N-1} \left\{ 2\sum_{l=1}^k \cos\left(2\pi \frac{lm}{N}\right) + 1 \right\} e^{2\pi i \frac{nm}{N}}\nonumber\\
 &=  \sum\limits_{l=-k}^k \frac{1}{N} \sum\limits_{m=0}^{N-1} e^{2\pi i \frac{ (n-l) m}{N}} = \sum\limits_{l=-k}^k \delta_{nl}\;,
\end{align}
which is the periodic step sequence
\begin{align}
 \zeta_{k,n} = \begin{cases}
            1 & \text{if}\quad n\!\!\!\mod\!N \leq k \quad\text{or}\quad n\!\!\!\mod\!N \geq N-k\\
            0 & \text{otherwise}
           \end{cases}\;.
\end{align}

We iteratively define the sequence $\zeta_k^{\ast l}$ of sequences
\begin{align}
\begin{aligned}
 \zeta_k^{\ast l} &:= (\zeta_k \ast\zeta_k^{\ast (l-1)})\;, & \zeta_k^{\ast1} &= \zeta_k\;.
\end{aligned}
\end{align}

Thus, the inverse Fourier transform $\varkappa = \mathcal{F}^{-1}[\,s\,]$ yields
\begin{align}
  \varkappa_n &= \mathcal{F}^{-1}[\,s\,]_n = \frac{1}{N}\sum_{m=0}^{N-1} s_m e^{2\pi i\frac{nm}{N}}\nonumber\\
  &= \frac{\gamma^2}{\alpha(\frac{2}{\alpha}+2k+1)}\sum_{l=0}^\infty \frac{1}{N}\sum_{m=0}^{N-1} \left(\frac{z_{k,m}}{\frac{2}{\alpha}+2k+1}\right)^le^{2\pi i\frac{nm}{N}}\nonumber\\
  &= \frac{\gamma^2}{\alpha(\frac{2}{\alpha}+2k+1)}\left\{\delta_{0n}+\sum_{l=1}^\infty \frac{\mathcal{F}^{-1}[\,z_{k}^l\,]_n}{(\frac{2}{\alpha}+2k+1)^l}\right\}\nonumber\\
              &= \frac{\gamma^2}{\alpha(\frac{2}{\alpha}+2k+1)}\left\{\delta_{0n}+\sum_{l=1}^\infty \frac{\zeta_{k,n}^{\ast l}}{(\frac{2}{\alpha}+2k+1)^l}\right\}\label{eq:varkappazeta}
\end{align}
Hence, the covariance $\varkappa_n$ between two nodes $i$ and $(i+n)$ is an infinite weighted sum of simple sequences.

\subsubsection{Monotonicity of $\zeta_k^{\ast l}$}\label{apdx: ring convolutions}
Let $\zeta_k$ be the periodic step sequence
\begin{align}
 \zeta_{k,n} = \begin{cases}
            1 & \text{if}\quad n\!\!\!\mod\!N \leq k \quad\text{or}\quad n\!\!\!\mod\!N \geq N-k\\
            0 & \text{otherwise}
           \end{cases}\;.
\end{align}
and let the sequence of sequences $\zeta_k^{\ast l}$ be defined by
\begin{align}
\begin{aligned}
 \zeta_k^{\ast l} &:= (\zeta_k \ast\zeta_k^{\ast (l-1)})\;, & \zeta_k^{\ast1} &= \zeta_k\;.
\end{aligned}
\end{align}
Furthermore, let $k,N\in \mathbb{N}$ and $\delta>0$ with $2k+1<N$.\\
We note that $\zeta_k^{\ast1}\,\widehat{=}\,\zeta_k$ is symmetric (i.e.
invariant under $n\!\!\mapsto\!\!-n$). Then, by induction, we find that, for all
$l$, $\zeta_k^{\ast l}$ is symmetric:
\begin{align}
  \zeta_{k,-n'}^{\ast l}=\zeta_{k,n'}^{\ast l}
\end{align}
More importantly, we note that, again by induction, for all $l$,
$\zeta_k^{\ast l}$ is monotonically decreasing in the interval
$n\in[0,\frac{N}{2})$, i.e.
\begin{align}
  \zeta_{k,n}^{\ast l} - \zeta_{k,n+1}^{\ast l}\geq0\;.
\end{align}

Since the sequence $\varkappa$ is a sum of sequences that are symmetric and monotonically
decreasing in the interval $n\in[0,\frac{N}{2})$ (compare \eqref{eq:varkappazeta}), we thus conclude that $\varkappa$ itself has these properties.

\subsection{The Difference $\varkappa_k-\varkappa_{k+1}$}\label{apdx: ring difference}

Equation \eqref{eq:sequencesum} yields the difference $\varkappa_k-\varkappa_{k+1}$:
\begin{align}
 \sum_{l=1}^k \left(\varkappa_{k-l}-\varkappa_{k+1-l}\right) - &2(\frac{1}{\alpha}+k)\left(\varkappa_{k}-\varkappa_{k+1}\right)+\sum_{l=1}^k \left(\varkappa_{k+l}-\varkappa_{k+1+l}\right) = 0\\
 \Rightarrow\quad \varkappa_0 - \varkappa_k - &2(\frac{1}{\alpha}+k)(\varkappa_{k} - \varkappa_{k+1}) + \varkappa_{k+1} - \varkappa_{2k+1} = 0\\
 \Rightarrow\quad \varkappa_{k} - &\varkappa_{k+1} = \frac{1}{\frac{2}{\alpha}+2k+1}\left(\varkappa_0-\varkappa_{2k+1}\right)
\end{align}

Since $\varkappa$ is monotonically decreasing in the interval $n\in[0,\frac{N}{2})$ for $2k+1<N$, $\varkappa_0>\varkappa_n$.
Importantly, $\varkappa_{2k+1}\neq\varkappa_0$ since we chose $k$ such that it fulfills $2k+1<N$.
Hence,
\begin{align}
 \varkappa_0-\varkappa_{2k+1} > 0 \quad\Rightarrow\quad \varkappa_k-\varkappa_{k+1}>0\;.
\end{align}

\subsection{Conclusion}
\label{sec:proofringconclusion}

$\varkappa_n$ is monotonically decreasing for $|n|<\frac{N}{2}$ and the farthermost connected unit is more correlated than the closest connected unit. Hence, connected units are strictly more correlated than unconnected units. Thus, $k$-ring topologies of this model are always reconstructible. 

\section{Evaluation of Reconstruction Errors}

Receiver operator characteristic (short: ROC or ROC curve) provide a method to
visualize and evaluate the quality of binary classifiers. In the manuscript, we
use ROC curves to evaluate the discriminative power of correlation thresholding as
classifier between links and non-links.

ROC curves and their usefulness to compare classifier properties are
discussed extensively in the literature (e.g., compare \cite{fawcett2004roc}).
For those who are not familiar with the concept we summarize the necessary information regarding our manuscript.\\

A binary classifier is a functions $h$ which classifies whether a sample $v\in\mathcal{M}$ belongs to a certain class ($h(v)=\text{True}$) or not ($h(v)=\text{False}$). $\mathcal{M}$ is called sample space.
\begin{align}
 h:\;\;\mathcal{M}&\rightarrow\{\text{False},\text{True}\}
\end{align}
Let $\mathcal{M}^+\subseteq\mathcal{M}$ be the set of samples actually belonging to class and let $\mathcal{M}^-\subseteq\mathcal{M}$ be a set of samples not belonging to that class. Let them have cardinalities $N^+:=|\mathcal{M}^+|$ and $N^-:=|\mathcal{M}^-|$, so that $\mathcal{M}=\mathcal{M}^+\cup\mathcal{M}^-$ and $N:=|\mathcal{M}|=N^++N^-$. Then a perfect classifier has to fulfill the conditions
\begin{align}
 v\in\mathcal{M}^+\;&\Leftrightarrow\;h(v)=\text{True}\\
 v\in\mathcal{M}^-\;&\Leftrightarrow\;h(v)=\text{False}\;.
\end{align}
However, real classifiers are usually imperfect; they produce false classifications.\\
These failures can either be false positive, if a sample is incorrectly classified as a member of the class, or false negative, if a member of the class is not identified as such. Correctly categorized samples constitute true positive or true negative classifications accordingly.\\
Let $\mathcal{T}^+,\mathcal{T}^-,\mathcal{F}^+,\mathcal{F}^-\subseteq\mathcal{M}$  be the subsets of true positive, true negative, false positive and false negative classifications. Hence,
\begin{align}
 \mathcal{T}^+\cup\mathcal{F}^-&=\mathcal{M}^+\\
 \mathcal{T}^-\cup\mathcal{F}^+&=\mathcal{M}^-\;.
\end{align}
The fraction of true positive classifications with respect to the overall
numbers of positive samples is called true positive rate
$t^+=\frac{|\mathcal{T}^+|}{|\mathcal{M}^+|}$ or \emph{sensitivity} and
$f^-=\frac{|\mathcal{F}^-|}{|\mathcal{M}^+|}$ is called false negative rate.
True negative rate or \emph{specificity} $t^-$ and the false positive rate $f^+$
are defined analogously.\\

Every non-trivial classifier depends on parameters which determine its output.
In the manuscript, classifiers depend on one criterion: the
correlation threshold. By varying this threshold and measuring sensitivity and specificity, a finger print of performance in $f^+$-$t^+$ space is obtained. This finger print is called ROC curve. \\
Depending on the shape of the curve the quality of the classifier can be extracted visually.\\
For example, consider the witless random classifier which decides at random with a probability $p$ if a sample is classified positively. For large $N^+$ the true positive rate is then $t^+\approx\frac{p\cdot N^+}{N^+}=p$. Same holds for the false positive rate in case of large $N^-$ since $f^+\approx\frac{p\cdot N^-}{N^-}=p$. Hence, $t^+=f^+$.\\
This is why the ROC of every random classifier lies on the identity in $f^+$-$t^+$ space.\\
The ROC curve of an ideal classifier has to intersect the point $(0,1)$ in $f^+$-$t^+$ space because no false positives and false negatives are produced for some criterion value.

When separating two classes by thresholding of a criterion value, the curve start at $(0,0)$ and end at $(1,1)$. If both sets can be separated, the classifier is perfect and the ROC has a rectangular shape. The area under the curve will be exactly $\textrm{AUC}=1$. Otherwise the integral will lead to smaller values.

For each network realization, we computed the correlation matrix $C$ and
employed a sliding threshold $\theta$ to reconstruct undirected network
representations $A'$ in the way discussed above. Plotting the true positive rate
$r_t(\theta)$ (the percentage of correctly inferred links) versus the false
positive rate $r_f(\theta)$ (the percentage of non-links that where erroneously
classified as links) results in the receiver-operator characteristic (ROC) of
the decision problem. The area under the curve $\AUC=\int r_t\;\dd r_f$ is a
benchmark for the evaluation of classifiers like discussed above.

 

\end{document}